\documentclass[%
 reprint,
 amsmath,amssymb,
 aps,
]{revtex4-1}

\usepackage{graphicx}% Include figure files
\usepackage{dcolumn}% Align table columns on decimal point
\usepackage{bm}% bold math
\usepackage{xcolor}
\usepackage{hyperref}% add hypertext capabilities
\usepackage[mathlines]{lineno}% Enable numbering of text and display math
\usepackage{natbib}
\begin{document}

\preprint{APS/123-QED}

\title{Multiscale Computational Modeling of Biofilm}% Force line breaks with \\
\thanks{A minireview}%

\author{Hanfeng Zhai}
%  \altaffiliation[Also at ]{Physics Department, XYZ University.}%Lines break automatically or can be forced with \\
 \email{hz253@cornell.edu}
\affiliation{\small
 Sibley School of Mechanical and Aerospace Engineering,\\
 Cornell University, Ithaca, NY 14850, USA
}%

% \collaboration{MUSO Collaboration}%\noaffiliation

%  \homepage{http://www.Second.institution.edu/~Charlie.Author}
% \affiliation{
%  Second institution and/or address\\
%  This line break forced% with \\
% }%
% \affiliation{
%  Third institution, the second for Charlie Author
% \affiliation{%
%  Authors' institution and/or address\\
%  This line break forced with \textbackslash\textbackslash
% }%

% \collaboration{CLEO Collaboration}%\noaffiliation

\date{\today}% It is always \today, today,

\begin{abstract}
We review the computation models for biofilm and bacteria cells, providing perspectives on biofilm's various properties and potential serving as engineering living materials (ELMs), considering the omnipresence of such biological matter. The minireview starts from the molecular regime, bottom-up to the mesoscale, to continuum, with an emphasis on the mesoscale algorithms such as dissipative particles dynamics (DPD) and individual-based modeling (IbM). Some representative works are highlighted considering different modeling methods on each scale. The advantages and limitations of each algorithm for different scales are elaborated given the existed research works. Specifically, the potential for IbM, also known as the discrete element method (DEM) is targeted for its accurate description of both biological and mechanical properties. 
\begin{description}
\item[Keywords]
Biomaterials; biofilms; multiscale computational modeling
% You may use the \texttt{description} environment to structure your abstract;
% use the optional argument of the \verb+\item+ command to give the category of each item. 
\end{description}
\end{abstract}

\maketitle

%\tableofcontents
\section{\label{intro}Introduction}
\begin{figure}[htbp]
    \centering
    \includegraphics[scale=0.19]{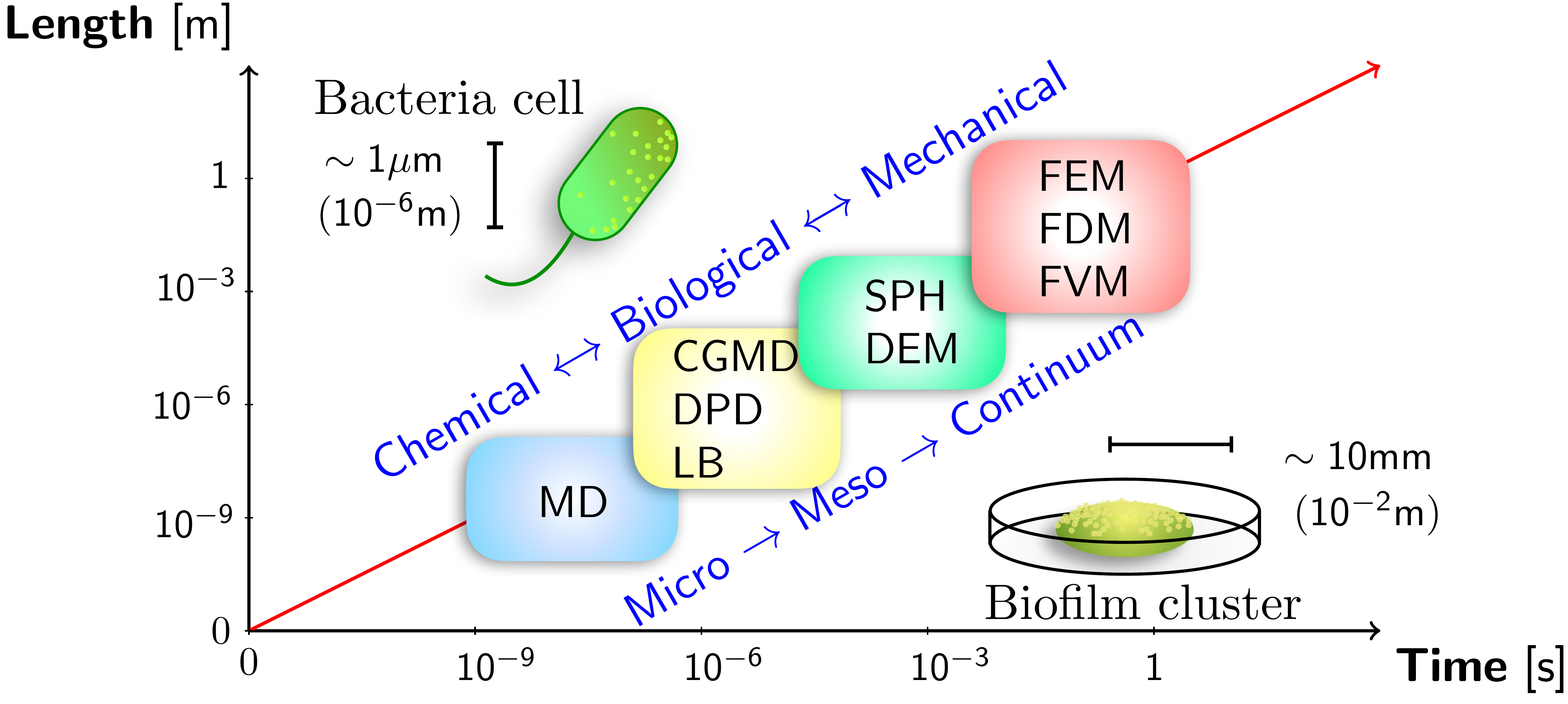}
    \caption{Schematic for multiscale computational methods for modeling biofilm, starting from the molecular scale, mesoscale, to continuum scale, where MD, DPD, etc., are abbreviations of different computation algorithms to be illustrated in the text. Note that the schematic on the left up corner indicates a single bacteria cell is approximately $1\mu \rm m$. The schematic on the right down corner indicates shows a matured biofilm "pancake" approximately $10 \rm mm$.}
    \label{biofilm_multiscale}
\end{figure}

Biofilms are bacteria communities adhered to surfaces that accommodate and clustered, exhibiting multiscale biomechanical behaviors \cite{colorado}. In this minireview, we review the multiscale computational modeling methods for simulating biofilms, with a concentration on mechanical properties and their applications or potentials for serving as engineering living materials (ELMs). The computational review regarding multiscale will follow Figure \ref{biofilm_multiscale}: we review all-atomic molecular dynamics (MD) techniques for molecular modeling biofilm regarding mostly on chemical and biological mechanisms, followed by a mesoscopic approach on dissipative particles dynamics (DPD) and lattice Boltzmann works; with further bottom-up to smoothed particles hydrodynamics (SPH) and discrete element method (DEM) modeling of biofilm. Note that DEM can also be called individual-based modeling (IbM) \& agent-based modeling (ABM), which is one of the most adopted methods for biofilm modeling due to its accurate description and coupling of biological processes, chemical reactions, and mechanical properties, as one of the most promising techniques for bridging multiscale multiphysics properties. 

\section{Molecular scale}

Molecular dynamics (MD) is a computer simulation technique that computes the atomic and molecular interactions pertaining time-based on Newton's law \cite{md}. First introduced and employed for simulating water in the 1950s and 1960s \cite{md_1, md_2}, followed by subsequent applications on biomolecular systems, such as protein or nucleic acids in the 1970s \cite{md, md_protein}. Traditionally, MD are classified as ab initio (first principle) MD (AIMD) and empirical MD, which differs from atomic forces calculation accounting for the potential fields where AIMD computes potential fields from quantum-mechanical calculations yet empirical MD assumes a prescribed field. Targeting biofilm approaches MD stands for empirical methods since it allows computation of larger scales. Since in MD applications atoms are ranged in the scale of $10^{-10} \sim 10^{-9}$m, in which biofilm clusters are hardly simulated limited by computational resources, most works adopting such methods were concentrated on the chemical and biological perspective. For instance, MD simulations can explain the mechanism of hydrogen bonds in the forming of polysaccharide Granulan, a gel forming matrix component of granular microbial biofilms \cite{md_biofilm_1}. Also, it can assist experiments to unveil the interactions between DNA and related ions of studying mucoid {\em Pseudomonas aeruginosa} biofilm \cite{md_biofilm_2}.
It can also be applied to study the membranes' interactions with Antimicrobial peptides (AMPs), for stable membrane binding \cite{md_biofilm_3}. In brief, the MD approach mainly tackles chemical and biological properties as unveiling the molecular insights of biofilm studies. Limited by computing powers, MD hardly unveils biofilms mechanical properties, in the scale of $10^{-9} \sim 10^{-6}$m.

% \footnote{Not sure if this explanation is of importance to an overview of MD}
% \begin{figure}
%     \centering
%     \includegraphics{biofilm_multiscale.pdf}
%     \caption{Microscopic pictures used to illustrate the multiscale nature of biofilm. The scale from top to down are illustrated from {\bf A} to {\bf E}. {\bf A}. the biofilm cultured in Petri dish, in the scale $10^{-2}$ to $10^{-1}$ m. {\bf B}. the cluster of bacteria cells, in the scale of $10^{-6}$ to $10^{-5}$m. {\bf C}. snapshot of a single bacteria cell, approximately $2 \sim 5 \mu$m. {\bf D}. zoomed view of EPS matrix, in the scale of $10^{-8}$ to $10^{-7}$m. {\bf E}. a schematic illustration of cell membrane, in the scale of $10^{-9}$ to $10^{-8}$m. Note that subfigures {\bf A}, {\bf C}, {\bf D} are acquired and modified from \url{https://www.quantamagazine.org/the-beautiful-intelligence-of-bacteria-and-other-microbes-20171113/}, Copyright \copyright Quanta Magazine. Subfigure {\bf C} is acquired from \url{https://www.nsf.gov/news/mmg/mmg_disp.jsp?med_id=71359&from=}, Copyright \copyright National Science Foundation. Subfigure {\bf E} is acquired from \url{https://pixels.com/featured/cell-membrane-artwork-pasieka.html}, Copyright \copyright Pixels \& Pasieka, 2013.}
%     \label{picture_multiscale}
% \end{figure}
% Within the molecular scale, the Hybrid Monte Carlo (MC) is also a celebrated method for molecular simulation \cite{mc_1}. Unlike MD, MC do not present information of time evolution but rather present configuration probabilities \cite{mc_rev}. MC has been employed 

\section{Mesoscale}

The mesoscale is defined between the molecular scale and the continuum scale, usually identified ranging from $10^{-6}\sim10^{-3}$m, where most mechanical and biological behavior of biofilm can be characterized in such a scale, as illustrated in Figure \ref{biofilm_multiscale}. Hence, we attach greater importance to such a regime. Following MD, coarse-grained MD (CGMD) is a method to use the simplified representation of a system for simulating its behavior, which is widely adopted in biochemical and biomolecular systems \cite{cgmd}, specifically for studying the biochemical properties of bacteria cells \cite{cgmd_bac_1, cgmd_bac_2, cgmd_bac_3}. CGMD is mostly adopted to simulate the system in the scale of micrometers, approximately the scale of a single bacteria cell, as we visualized in Figure \ref{biofilm_multiscale}. As an extension to MD, CGMD simulations was still limited in scales constrained by computational power, as it can be employed to explain the chemical and biological essence of biofilm dewetting phenomenon of {\em Bacillus subtilis} \cite{cgmd_1}, cross-validating experiments proposed theory of biological transporation \cite{cgmd_2}, and providing molecular insights for developing anti-bacteria silver drugs \cite{cgmd_3}.

% Hence, the behavior of single bacteria and interactions between each are mostly simulated using such a method. For example, the biofilm formed by Bacillus subtilis exhibits non-wetting behavior that can be attributed to layers formed by the protein BslA. CGMD simulations unveiled that specific BslA types can be identified as Janus particles, due to its orientation difference at the interface
On top of CGMD \cite{dpd_cg}, another method that bottoms from MD computation strive to accurately depict the mesoscale is Dissipative Particle Dynamics (DPD). DPD is a stochastic simulation that is widely applied for complex fluids, initially proposed by Hoogerbrugge and Koelman \cite{dpd_1, dpd_2}, assigning statistical mechanics information beads that conserve chemical and physical properties, has mostly applied to microfluidics and complex fluid modeling. Has already been widely applied to small scales in biological systems such as cell membranes and lipid bilayers \cite{dpd_pre_1, dpd_pre_2, dpd_pre_3, dpd_pre_4}, DPD was initially applied to model biofilm in 2011 by Xu et al. \cite{dpd_biofilm_2}, where the fluid flow interactions and transport phenomena were keenly focused. Bacteria cells can also be modeled as a combination of hundreds of DPD beads to model biofilms \cite{dpd_biofilm_1}. What's more, DPD was adopted to investigate the biofilm constitutive model \cite{dpd_biofilm_3} and design of antibiotic drug design \cite{dpd_biofilm_4}. Lattice Boltzmann method (LBM) is a particle-based, bottom-up model that can be employed for tracing the dynamics and properties of individual bacterial cells \cite{lb_1}, originated from classical statistical physics and lattice gas automata \cite{lb_intro}. Applications of LBM for biofilm modeling dated back to 1999 \cite{lb_2, lb_4}, which sparks a series of study on applying LBM for biofilm growth simulation in the 2000s \cite{lb_2000_1, lb_2000_2, lb_2000_3, lb_2000_4, lb_2000_5}. 

\begin{figure}
    \centering
    \includegraphics[scale=0.5]{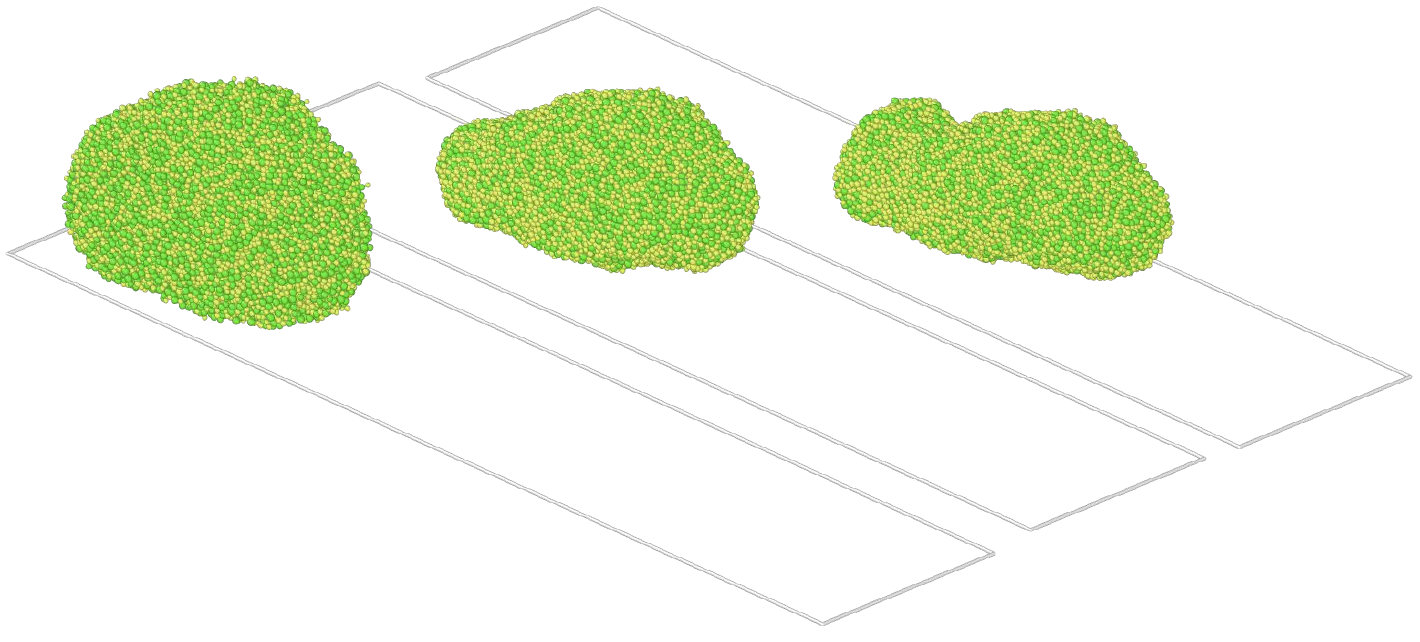}
    \caption{Simulation snapshots of biofilm cluster under shear flow generated from NUFEB \cite{ibm_9, ibm_12} implemented in LAMMPS \cite{lammps}.}
    \label{snapshots}
\end{figure}

Specifically, the most widely adopted method to model biofilms is the discrete element method (DEM), wherein biofilm modeling they are mostly referred to as individual-based modeling (IbM) or agent-based modeling (AbM). For DEM singular bacteria cells are usually identified as single elements and the duplications and interactions are based on different biological mathematical models. It is particularly suitable for biofilm modeling due to its successful coupling and accurate description of hydrodynamics, thermodynamics, biological processes, etc., and can act as a bridge between behavior at the individual and community levels \cite{ibm_rev_2}. In the past few years, a surge of DEM-based biofilm numerical models occurred tackle different questions in biofilms \cite{ibm_rev}. Among the numerous biofilm models, most proposed DEM biofilm models can couple physical and biological processes \cite{ibm_1, ibm_2, ibm_3, ibm_4, ibm_5, ibm_8, ibm_9, ibm_10, ibm_11, ibm_12, ibm_13, ibm_14, ibm_15, ibm_16}, amongst some cannot include the metabolic, as to quantify the biofilm interactions with the external environment \cite{ibm_1, ibm_13, ibm_15}. One of the main motivations of biofilm studies is bacteria communities generate extracellular polymeric substances (EPS) as an external engineering matrix to adhere and protect cells from drugs and environmental changes. Several proposed DEM models can successfully model EPS \cite{ibm_6, ibm_7, ibm_9, ibm_10, ibm_11, ibm_12, ibm_16}, which can be adopted as potential tools for studying the EPS mechanism to utilize biofilms as ELMs.

Another simulation technique widely applied in the mesoscale utilizing particle-based method yet for computing the mechanics of continuum media is smoothed particles hydrodynamics (SPH), first proposed by Gingold and Monaghan \cite{sph_base_1} and Lucy \cite{sph_base_2} in the field of aerospace. Similar to the DEM schemes, SPH methods can also successfully represent both bioreaction and nutrient diffusion with also accounting for deformation and interface erosion, according to Soleimani et al. \cite{sph_biofilm_1}, where the difference is SPH is based on a continuum approach. SPH can also be applied to study the mechanics of EPS \cite{sph_biofilm_2} and chemotaxis \cite{sph_biofilm_3} for biofilm under fluid flow-induced deformation.

To summary, the mesoscale computational methods provide decent characterizations of the multiphysics nature of biofilm. Coarse-grained molecular like CGMD and DPD models proffer accurate depictions of chemotaxis phenomena while DEM offers a satisfactory bridge between meso to continuum for coupling mechanical, biological and even chemoaxis biofilm signature, ranging in the scale from $10^{-6}$ to $10^{-3}$m. 

\section{Continuum scale}

In the continuum regime, most numerical methods aim to discretize ordinary or partial differential equations that govern the mechanical, chemical, and biological process, whereas commonly employed methods include finite element method (FEM), finite difference method (FDM), finite volume method (FVM), etc. Especially, the computational modeling provides insights on biofilms mechanical properties validating experiments since most biomechanical tests are conducted on the scale of $10^{-3} \sim10^{-1}$m. As one of the most adopted computational mechanics methods, FEM subdivides the computational domain into smaller subdomains called finite elements, achieved through the construction of meshing, which can be traced back to the 1940s by Hrennikoff \cite{fem_rev_1} and Courant \cite{fem_rev_2}. The extended FEM (XFEM) method can study boundary layer behavior in elliptic equations, which can be further applied to linearized biofilm growth \cite{fem_2}. Followed up, XFEM can either be combined with diffusion-reaction and show the relation between colony morphology and nutrient deletion \cite{fem_1}; or with the level set method, algorithms widely used in multiphase flow computation, for simulate biofilms growth \cite{fem_5}. When investigating biofilms detachment under fluid flow with FEM, the fluid-structure interactions are of importance \cite{fem_3}. Notably, Feng et al. incorporate the time-discontinuous Galerkin (TDG) method as solution strategies for a multi-dimensional multi-species biofilm growth model \cite{fem_4}. In short, multiphysics-combined XFEM methods accurately describe biofilms' mechanical behavior as a continuum approach to computing biofilms on the scale of $10^{-3} \sim 10^{-2}$m.

% These methods are particularly suitable for validations in biofilms experiments.

Not as widely adopted as FEM, FVM biofilm modeling begins in 1993 by combining the FVM scheme with tracking of the time
evolution of the interface \cite{fvm_1993}. Followed FVM schemes by Zurek's group employs implicit Eulerian solver attempts to quantify diffusion and biomass fraction into fluid dynamics \cite{fvm_1, fvm_2}. Taking advantage of its simpleness, FDM were also widely employed for coupling diffusion, growth, biomass and nutrients concentrations in both 2D and 3D \cite{fdm_1, fdm_2, fdm_3, fdm_4, fdm_5, fdm_6}. In fine, continuum models strives to couple biological signature with growth and physics of biofilm models solved by discretizing differential equations, offers good characterizations for the overall biofilm behavior in the scale of $10^{-3}\sim 10^{-1}$m.

\section{Conclusion}

Different numerical methods satisfied for different scales are briefly explained and their applications to issues involved in biofilms modeling are reviewed. On the molecular scale, the chemical and biological properties can be decently modeled with MD. In the mesoscale, CGMD and DPD offers also good characterizations of biochemical properties with a larger scale compared with MD; while DEM provides fruitful information ranging from chemical to mechanical, as the most adopted modeling method for biofilm. SPH bottoms up as closer to the continuum regime yet not as wide applied. Continuum modeling methods such as FEM and FDM can also successfully characterize growth, biomass diffusion, with also mechanical signature, yet hardly gives any chemical information due to the nature of the model. As specifically for the application of ELMs, since we are keenly focused on mechanical aspects with curious on biological mechanism; DEM can be adopted as one of the best tool for: (1) the abundant of existed model; (2) the multiscale nature of the model; (3) it can easily bridged to continuum regime, and offers more multiphysics information compared with continuum models.

\bibliographystyle{aipauth4-1}
\bibliography{ref}
\end{document}